\documentclass[onecolumn,showpacs,11pt]{revtex4}
\usepackage{axodraw}
\usepackage{graphicx}
\usepackage{dcolumn}
\usepackage{bm}
\begin{document}
\newcommand{\hs}{\hspace*{0.5cm}}
\newcommand{\vs}{\vspace*{0.5cm}}
\newcommand{\be}{\begin{equation}}
\newcommand{\ee}{\end{equation}}
\newcommand{\bea}{\begin{eqnarray}}
\newcommand{\eea}{\end{eqnarray}}
\newcommand{\ben}{\begin{enumerate}}
\newcommand{\een}{\end{enumerate}}
\newcommand{\bde}{\begin{widetext}}
\newcommand{\ede}{\end{widetext}}
\newcommand{\nn}{\nonumber}
\newcommand{\crn}{\nonumber \\}
\newcommand{\Tr}{\mathrm{Tr}}
\newcommand{\non}{\nonumber}
\newcommand{\noi}{\noindent}
\newcommand{\al}{\alpha}
\newcommand{\la}{\lambda}
\newcommand{\bet}{\beta}
\newcommand{\ga}{\gamma}
\newcommand{\va}{\varphi}
\newcommand{\om}{\omega}
\newcommand{\pa}{\partial}
\newcommand{\+}{\dagger}
\newcommand{\fr}{\frac}
\newcommand{\bc}{\begin{center}}
\newcommand{\ec}{\end{center}}
\newcommand{\Ga}{\Gamma}
\newcommand{\de}{\delta}
\newcommand{\De}{\Delta}
\newcommand{\ep}{\epsilon}
\newcommand{\varep}{\varepsilon}
\newcommand{\ka}{\kappa}
\newcommand{\La}{\Lambda}
\newcommand{\si}{\sigma}
\newcommand{\Si}{\Sigma}
\newcommand{\ta}{\tau}
\newcommand{\up}{\upsilon}
\newcommand{\Up}{\Upsilon}
\newcommand{\ze}{\zeta}
\newcommand{\ps}{\psi}
\newcommand{\Ps}{\Psi}
\newcommand{\ph}{\phi}
\newcommand{\vph}{\varphi}
\newcommand{\Ph}{\Phi}
\newcommand{\Om}{\Omega}

\title{The 3-3-1 model with $S_4$ flavor symmetry}

\author{P. V. Dong}
\email {pvdong@iop.vast.ac.vn} \affiliation{Institute of Physics,
VAST, P. O. Box 429, Bo Ho, Hanoi 10000, Vietnam}
\author{H. N. Long}
\email{hnlong@iop.vast.ac.vn} \affiliation{Institute of Physics,
VAST, P. O. Box 429, Bo Ho, Hanoi 10000, Vietnam}
\author{D. V. Soa}
\affiliation{Department of Physics, Hanoi University of Education,
Hanoi, Vietnam}
\author{V. V. Vien}
\affiliation{Department of Physics, Tay-Nguyen University,
Tay-Nguyen, Vietnam}

\date{\today}

\begin{abstract}

We construct a 3-3-1 model based on family symmetry $S_4$
responsible for the neutrino and quark masses. The tribimaximal
neutrino mixing and the diagonal quark mixing have been obtained.
The new lepton charge $\mathcal{L}$ related to the ordinary lepton
charge $L$ and a $\mathrm{SU}(3)$ charge by
$L=\fr{2}{\sqrt{3}}T_8+\mathcal{L}$ and the lepton parity
$P_l=(-)^L$ known as a residual symmetry of $L$ have been
introduced which provide insights in this kind of model. The
expected vacuum alignments resulting in potential minimization can
origin from appropriate violation terms of $S_4$ and
$\mathcal{L}$. The smallness of seesaw contributions can be
explained from the existence of such terms too. If $P_l$ is not
broken by the vacuum values of the scalar fields, there is no
mixing between the exotic
 and the ordinary quarks at the
tree level.

\end{abstract}


\maketitle

\section{\label{intro}Introduction}

In the standard model the fundamental fermions come in families.
In writing down the theory one may start by first introducing just
one family, then one may repeat the same procedure by introducing
copies of the first family. Why do quarks and leptons come in
repetitive structures--families? How many families are there? How
to understand the interrelation and mass-hierarchy between the
families? In addition, the standard model cannot explain the tiny
masses and mixing profile of neutrinos, and the close-to-unity of
quark mixing matrix as well \cite{pdg}. These have been the
central puzzles known as the flavor question in particle physics
beyond the standard model.

The current neutrino experimental data are consistent with the
tribimaximal form proposed by Harrison-Perkins-Scott (HPS), which
apart from the phase redefinitions, is given by \cite{hps}
\begin{eqnarray}
U_{\mathrm{HPS}}=\left(
\begin{array}{ccc}
\frac{2}{\sqrt{6}}       &\frac{1}{\sqrt{3}}  &0\\
-\frac{1}{\sqrt{6}}      &\frac{1}{\sqrt{3}}  &\frac{1}{\sqrt{2}}\\
-\frac{1}{\sqrt{6}}      &\frac{1}{\sqrt{3}}  &-\frac{1}{\sqrt{2}}
\end{array}\right),\label{eq:1}
\end{eqnarray}
where the large mixing angles are completely different from the
quark mixing ones defined by the Cabibbo-Kobayashi-Maskawa (CKM)
matrix. It is an interesting challenge to formulate dynamical
principles that can lead to the flavor mixing patterns for quarks
and leptons given in a completely natural way as first
approximations. A fascinating way seems to be the use of some
discrete non-Abelian groups \cite{kj} as family symmetries added
to the standard model gauge group. There is a series of models
based on the group $A_4$ \cite{A4,dlsh}, $T'$ \cite{T'}, and more
recently $S_4$ \cite{old S4, new S4}---the group of permutations
of four objects, which is also the symmetry group of the cube.

We would like to extend the above application to the
$\mathrm{SU}(3)_C\otimes \mathrm{SU}(3)_L \otimes \mathrm{U}(1)_X$
(3-3-1) gauge model \cite{331m,331r,ecn331} because of the
following. The $[\mathrm{SU}(3)_L]^3$ anomaly cancelation in the
model requires the number of $\mathrm{SU}(3)_L$ fermion triplets
to equal that of antitriplets. Taking into account an unrestricted
number of standard model families with corresponding extensions of
lepton and quark representations, the number of families results
in a multiple of 3. Furthermore the QCD asymptotic condition
constrains the number of quark families to be lesser than or equal
to 5. The family number is exact 3. The model thus provides a
partial explanation of the family number, as also required by
flavor symmetries such as $S_4$ for 3-dimensional representations.
In addition, due to the anomaly cancelation one family of quarks
has to transform under $\mathrm{SU}(3)_L$ differently from the two
others. We should look for a family symmetry group with 2- and
3-dimensional irreducible representations respectively acting on
the 2- and 3-family indices, the simplest of which is just $S_4$.
Note that $S_4$ has not been considered before in the kind of the
3-3-1 model. For the similar works on $A_4$, let us call the
reader's attention to Refs. \cite{dlsh}.

There are two typical variants of the 3-3-1 model as far as lepton
sectors are concerned. In the minimal version, three
$\mathrm{SU}(3)_L$ lepton triplets are of the form
$(\nu_L,l_L,l^c_R)$, where $l_{R}$ are ordinary right-handed
charged-leptons \cite{331m}. In the second version, the third
components of lepton triplets include right-handed neutrinos,
respectively, $(\nu_L,l_L,\nu^c_R)$ \cite{331r}. In trying to
recover the tribimaximal form in present work, by analysis a
possibility close to the typical versions is when we replace the
right-handed neutrinos by new standard model fermion singlets
($N_R$) with vanishing lepton-number \cite{matd}. The resulting
model is near that of our previous work in \cite{dlsh}. The
neutrinos thus gain masses from only contributions of SU(3)$_L$
scalar antisextets. The antisextets contain tiny vacuum
expectation values (VEVs) in the first components, similar to the
cases of the standard model with scalar triplets. To avoid the
decay of $Z$ into the Majorons associated with these components,
the lepton-number violating potential should be turned on. The
lepton charge is therefore no longer of an exact symmetry; thereby
the Majorons can get large enough masses to escape from the $Z$
decay \cite{matd}. Assuming the antisextets very heavy, the
potential minimization can provide a natural explanation of the
expected vacuum alignments as well as the smallness of seesaw
contributions responsible for neutrino mass.

The rest of this article is organized as follows. In Sec.
\ref{model}, we propose the model with $S_4$. The masses and
mixing matrices of leptons and quarks are obtained then. In Sec.
\ref{vev} we consider the Higgs potential and minimization
conditions. We summarize our results and make conclusions in
Sec.~\ref{conclus}. Appendix \ref{apa} is devoted to $S_4$ group
with its Clebsch-Gordan coefficients. Appendix \ref{apt} presents
the lepton numbers and lepton parities of model particles.

\section{\label{model}The model}

The fermions in this model under $[\mathrm{SU}(3)_L,
\mathrm{U}(1)_X, \mathrm{U}(1)_\mathcal{L},\underline{S}_4]$
symmetries, respectively, transform as \bea \psi_{L} &\equiv&
\psi_{1,2,3L}=\left(
  \begin{array}{c}
    \nu_{1,2,3L} \\
    l_{1,2,3L} \\
    N^c_{1,2,3R} \\
  \end{array}
\right)\sim [3,-1/3,2/3,\underline{3}],\\
l_{1R}&\sim&[1,-1,1,\underline{1}],\hs l_R\equiv l_{2,3R}\sim[1,-1,1,\underline{2}],\\
Q_{3L}&=& \left(
  \begin{array}{c}
    u_{3L} \\
    d_{3L} \\
    U_{L} \\
  \end{array}
\right)\sim[3,1/3,-1/3,\underline{1}],\hs Q_{L}\equiv Q_{1,2L}=
\left(
  \begin{array}{c}
    d_{1,2L} \\
    -u_{1,2L} \\
    D_{1,2L} \\
  \end{array}
\right)\sim[3^*,0,1/3,\underline{2}], \\
u_{R}&\equiv&u_{1,2,3R}\sim[1,2/3,0,\underline{3}],\hs d_{R}\equiv d_{1,2,3R}\sim[1,-1/3,0,\underline{3}],\\
U_R&\sim&[1,2/3,-1,\underline{1}],\hs D_R\equiv
D_{1,2R}\sim[1,-1/3,1,\underline{2}],\eea where the numbered
subscripts on field indicate to respective families which also in
order define components of their $S_4$ multiplet representation.
The reader can see in Appendix \ref{apa} for more details of the
$S_4$ group representations. As usual, the $X$ charge is related
to the electric charge operator as $Q=T_3-\fr{1}{\sqrt{3}}T_8+X$
where $T_a$ $(a=1,2,...,8)$ are $\mathrm{SU}(3)_L$ charges,
satisfying $\mathrm{Tr}[T_aT_b]=\fr 1 2 \de_{ab}$.

The $N_R$ as above mentioned are exotic neutral fermions having
the lepton number $L(N_R)=0$ \cite{matd,dlsh}. Hence the lepton
number $L$ in this model does not commute with the gauge symmetry.
We can therefore search for a new conserved charge $\mathcal{L}$
as given in the square brackets above, which is defined in terms
of the ordinary lepton number by
$L=\fr{2}{\sqrt{3}}T_8+\mathcal{L}$ \cite{clong,dlsh}. This
definition is only convenient one for accounting the global lepton
numbers of the model particles, because the $T_8$ is a gauged
charge, and thus $L$ consequently gauged. The gauging of the $L$
charge deserves further studies, where in the present work we will
take it globally. This is possible since the $T_8$ can be
considered as the charge of a group replication of
$\mathrm{SU}(3)_L$ but taken globally, thus $L$ is not gauged.
Finally, the lepton charge arranged in the way is to suppress
unwanted interactions (due to $\mathrm{U}(1)_{\mathcal{L}}$
symmetry) to yield the tribimaximal form as shown below. $U$ and
$D_{1,2}$ as supplied are exotic quarks carrying lepton numbers
$L(U)=-1$ and $L(D_{1,2})=1$, known as leptoquarks.

The lepton parity is introduced as follows $P_l=(-)^L$, which is a
residual symmetry of $L$. The particles possess $L=0,\pm 2$ such
as $N_R$, ordinary quarks and bileptons having $P_l=1$; the
particles with $L=\pm 1$ such as ordinary leptons and exotic
quarks have $P_l=-1$. Any non-zero VEV with odd parity, $P_l=-1$,
will break this symmetry spontaneously. For convenience in
reading, the numbers $L$ and $P_l$ of the component particles are
given in Appendix~\ref{apt}.

In the following, we consider possibilities of generating the
masses for the fermions. The scalar multiplets needed for the
purpose are introduced accordingly.

\section{fermion mass}

\subsection{Lepton mass}

To generate masses for the charged leptons, we need two scalar
multiplets:
\bea \phi = \left(%
\begin{array}{c}
  \phi^+_1 \\
  \phi^0_2 \\
  \phi^+_3 \\
\end{array}%
\right)\sim [3,2/3,-1/3, \underline{3}],\hs \phi' = \left(%
\begin{array}{c}
  \phi'^+_1 \\
  \phi'^0_2 \\
  \phi'^+_3 \\
\end{array}%
\right)\sim [3,2/3,-1/3, \underline{3}'], \eea with the VEVs
$\langle \phi \rangle = (v,v,v)$ and $\langle \phi' \rangle =
(v',v',v')$ written as those of $S_4$ components respectively
(these will be derived from the potential minimization
conditions). Here and after, the number subscripts on the
component scalar fields are indices of $\mathrm{SU}(3)_L$. The
$S_4$ indices are discarded and should be understood. The Yukawa
interactions are \bea -\mathcal{L}_{l}=h_1 (\bar{\psi}_L
\phi)_{\underline{1}} l_{1R}+h_2 (\bar{\psi}_L
\phi)_{\underline{2}} l_{R}+h_3 (\bar{\psi}_L
\phi')_{\underline{2}} l_{R}+h.c.\eea The mass Lagrangian of the
charged leptons reads
$-\mathcal{L}^{\mathrm{mass}}_l=(\bar{l}_{1L},\bar{l}_{2L},\bar{l}_{3L})
M_l (l_{1R},l_{2R},l_{3R})^T+h.c.$, \bea M_l=
\left(%
\begin{array}{ccc}
  h_1v & h_2v-h_3v' & h_2 v+h_3v' \\
   h_1v & (h_2v-h_3v')\om & (h_2 v+h_3v')\om^2 \\
  h_1v &  (h_2v-h_3v')\om^2 & (h_2 v+h_3v')\om \\
\end{array}%
\right).\eea The mass matrix is then diagonalized, \bea U^\dagger_L M_lU_R=\left(%
\begin{array}{ccc}
  \sqrt{3}h_1 v & 0 & 0 \\
  0 & \sqrt{3}(h_2 v - h_3v') & 0 \\
  0 & 0 & \sqrt{3}(h_2 v+h_3v') \\
\end{array}%
\right)=\left(%
\begin{array}{ccc}
  m_e & 0 & 0 \\
  0 & m_\mu & 0 \\
  0 & 0 & m_\tau \\
\end{array}%
\right),\eea where \bea U_L=\fr{1}{\sqrt{3}}\left(%
\begin{array}{ccc}
  1 & 1 & 1 \\
  1 & \om & \om^2 \\
  1 & \om^2 & \om \\
\end{array}%
\right),\hs U_R=1.\label{lep}\eea We see that the masses of muon
and tauon are separated by the $\phi'$ triplet. This is the reason
why we introduce $\phi'$ in addition to $\phi$.

Notice that the couplings $\bar{\psi}^c_L \psi_L\phi$ and
$\bar{\psi}^c_L \psi_L\phi'$ are suppressed because of the
$\mathcal{L}$--symmetry violation. Therefore
$\bar{\psi}^c_L\psi_L$ can couple to SU(3)$_L$ antisextets instead
to generate masses for the neutrinos. The antisextets in this
model transform as \bea \sigma=
\left(%
\begin{array}{ccc}
  \sigma^0_{11} & \sigma^+_{12} & \sigma^0_{13} \\
  \sigma^+_{12} & \sigma^{++}_{22} & \sigma^+_{23} \\
  \sigma^0_{13} & \sigma^+_{23} & \sigma^0_{33} \\
\end{array}%
\right)\sim [6^*,2/3,-4/3,\underline{1}], \eea \bea s=
\left(%
\begin{array}{ccc}
  s^0_{11} & s^+_{12} & s^0_{13} \\
  s^+_{12} & s^{++}_{22} & s^+_{23} \\
  s^0_{13} & s^+_{23} & s^0_{33} \\
\end{array}%
\right)\sim [6^*,2/3,-4/3,\underline{3}]. \eea The Yukawa
interactions are \bea -\mathcal{L}_\nu&=&\fr 1 2 x (\bar{\psi}^c_L
\psi_L)_{\underline{1}}\sigma+\fr 1 2 y (\bar{\psi}^c_L
\psi_L)_{\underline{3}}s+h.c.\crn &=& \fr 1 2
x(\bar{\psi}^c_{1L}\psi_{1L}+\bar{\psi}^c_{2L}\psi_{2L}+\bar{\psi}^c_{3L}\psi_{3L})\sigma\crn
&&
+y(\bar{\psi}^c_{2L}\psi_{3L}s_1+\bar{\psi}^c_{3L}\psi_{1L}s_2+\bar{\psi}^c_{1L}\psi_{2L}s_3)\crn
&&+h.c.\label{yn}\eea

The VEV of $s$ is set as $(\langle s_1\rangle,0,0)$ under $S_4$
(which is also a natural minimization condition for the scalar
potential), where \bea
\langle s_1\rangle=\left(%
\begin{array}{ccc}
  \la_{s} & 0 & v_{s} \\
  0 & 0 & 0 \\
  v_{s} & 0 & \Lambda_{s} \\
\end{array}%
\right).\label{s1}\eea The VEV of $\sigma$ is \bea
\langle \sigma \rangle=\left(%
\begin{array}{ccc}
  \la_\sigma & 0 & v_\sigma \\
  0 & 0 & 0 \\
  v_\sigma & 0 & \La_\sigma \\
\end{array}%
\right).\label{sim}\eea The mass Lagrangian for the neutrinos is
defined by \bea -\mathcal{L}^{\mathrm{mass}}_\nu=\fr 1 2
\bar{\chi}^c_L M_\nu \chi_L+ h.c.,\hs  \chi_L\equiv
\left(%
\begin{array}{c}
  \nu_L \\
  N^c_R \\
\end{array}%
\right),\hs M_\nu\equiv\left(%
\begin{array}{cc}
  M_L & M^T_D \\
  M_D & M_R \\
\end{array}%
\right),\label{nm}\eea where $\nu=(\nu_{1},\nu_{2},\nu_{3})^T$ and
$N=(N_1,N_2,N_3)^T$. The mass matrices are then obtained by
\bea M_{L,R,D}=\left(%
\begin{array}{ccc}
  a_{L,R,D} & 0 & 0 \\
  0 & a_{L,R,D} & b_{L,R,D} \\
  0 & b_{L,R,D} & a_{L,R,D} \\
\end{array}%
\right),\eea with \bea a_L=x \la_\sigma,\ a_D=x v_\sigma,\ a_R=x
\La_\sigma,\ b_L=y \la_{s},\ b_D= y v_s,\ b_R=y \La_s.\eea

The VEVs $\La_{\sigma,s}$ break the 3-3-1 gauge symmetry down to
that of the standard model, and provide the masses for the neutral
fermions $N_R$ and the new gauge bosons: the neutral $Z'$ and the
charged $Y^{\pm}$ and $X^{0,0*}$. The $\la_{\sigma,s}$ and
$v_{\sigma,s}$ belong to the second stage of the symmetry breaking
from the standard model down to the $\mathrm{SU}(3)_C \otimes
\mathrm{U}(1)_Q$ symmetry, and contribute the masses to the
neutrinos. Hence, to keep a consistency we assume that
$\La_{\sigma,s}\gg v_{\sigma,s},\la_{\sigma,s}$. The natural
smallness of the lepton number violating VEVs $\la_{\sigma,s}$ and
$v_{\sigma,s}$ will be explained in Section \ref{vev}. Three
active-neutrinos ($\sim\nu_L$) therefore gain masses via a
combination of type I and type II seesaw mechanisms derived from
(\ref{nm}) as \bea M_{\mathrm{eff}}=M_L-M_D^TM_R^{-1}M_D=
\left(%
\begin{array}{ccc}
  a' & 0 & 0 \\
  0 & a & b \\
  0 & b & a \\
\end{array}%
\right),\label{neu}\eea where \bea a'&=&a_L-\fr{a^2_D}{a_R},\crn
a&=&a_L+2a_Db_D\fr{b_R}{a_R^2-b^2_R}-(a^2_D+b^2_D)\fr{a_R}{a_R^2-b^2_R},\crn
b&=&b_L-2a_Db_D\fr{a_R}{a_R^2-b^2_R}+(a^2_D+b^2_D)\fr{b_R}{a_R^2-b^2_R}.\label{neu3}\eea

We can diagonalize the mass matrix (\ref{neu}) as follows: \bea
U^T_\nu M_{\mathrm{eff}} U_\nu=\left(%
\begin{array}{ccc}
  a+b & 0 & 0 \\
  0 & a' & 0 \\
  0 & 0 & b-a \\
\end{array}%
\right)=\left(%
\begin{array}{ccc}
  m_1 & 0 & 0 \\
  0 & m_2 & 0 \\
  0 & 0 & m_3 \\
\end{array}%
\right),\label{neu2} \eea where \bea U_\nu=\left(%
\begin{array}{ccc}
  0 & 1 & 0 \\
  \fr{1}{\sqrt{2}} & 0 & -\fr{1}{\sqrt{2}} \\
  \fr{1}{\sqrt{2}} & 0 & \fr{1}{\sqrt{2}} \\
\end{array}%
\right)\left(%
\begin{array}{ccc}
  1 & 0 & 0 \\
  0 & 1 & 0 \\
  0 & 0 & -i \\
\end{array}%
\right).\label{neu1}\eea Combined with (\ref{lep}), the lepton
mixing matrix yields the tribimaximal form: \bea U^\dagger_L
U_\nu=\left(%
\begin{array}{ccc}
  \sqrt{2/3} & 1/\sqrt{3} & 0 \\
  -1/\sqrt{6} & 1/\sqrt{3} & 1/\sqrt{2} \\
  -1/\sqrt{6} & 1/\sqrt{3} & -1/\sqrt{2} \\
\end{array}%
\right)=U_{\mathrm{HPS}},\eea which is a main result of the paper.

If the lepton parity is an exact and spontaneously unbroken
symmetry, the $a_D$ and $b_D$ vanish. The neutrinos then gain
masses only from the type II seesaw due to the VEVs of first
components of $\sigma$ and $s$, as we can see from (\ref{neu3})
with $a_D=b_D=0$. If this parity is broken, there is no reason to
prevent the 13 and 31 components of $\sigma$ and $s$ from getting
nonzero VEVs as given in $a_D$, $b_D$. The neutrino masses
therefore gain additional contributions from the type I seesaw as
well. Deviations from the tribimaximal form if required can be
further explained by $\mathrm{S}_4$ breaking soft-terms, or if
$\mathcal{L}$ was slightly violated, the terms breaking this
charge as mentioned would also give contributions.

\subsection{Quark mass}

To generate masses for quarks, we additionally acquire the
following scalar multiplets: \bea \chi&=&
\left(%
\begin{array}{c}
  \chi^0_1 \\
  \chi^-_2 \\
  \chi^0_3 \\
\end{array}%
\right)\sim[3,-1/3,2/3,\underline{1}],\\ \eta&=&
\left(%
\begin{array}{c}
  \eta^0_1 \\
  \eta^-_2 \\
  \eta^0_3 \\
\end{array}%
\right)\sim[3,-1/3,-1/3,\underline{3}],\hs \eta'=
\left(%
\begin{array}{c}
  \eta'^0_1 \\
  \eta'^-_2 \\
  \eta'^0_3 \\
\end{array}%
\right)\sim[3,-1/3,-1/3,\underline{3}'].\eea The Yukawa
interactions are \bea -\mathcal{L}_q &=& f_3 \bar{Q}_{3L}\chi U_R
+ f \bar{Q}_{L}\chi^* D_{R}\crn &&+h^d_3 \bar{Q}_{3L}(\phi
d_R)_{\underline{1}} + h^u_3 \bar{Q}_{3L}(\eta
u_R)_{\underline{1}}\crn && + h^u \bar{Q}_{L}(\phi^*
u_R)_{\underline{2}}+ h^d \bar{Q}_{L}(\eta^*
d_R)_{\underline{2}}\crn &&+h'^u \bar{Q}_{L}(\phi'^*
u_R)_{\underline{2}}+h'^d \bar{Q}_{L}(\eta'^*
d_R)_{\underline{2}}\crn &&+h.c.\label{yv}\eea

Suppose that the VEVs of $\eta$, $\eta'$ and $\chi$ are $(u,u,u)$,
$(u',u',u')$ and $w$, where $u=\langle \eta^0_1\rangle$,
$u'=\langle \eta'^0_1\rangle$, $w=\langle \chi^0_3\rangle$. The
other VEVs $\langle \eta^0_3\rangle$, $\langle \eta'^0_3\rangle$,
$\langle\chi^0_1\rangle$ vanish if the lepton parity is conserved.
Otherwise they can develop VEVs. In addition, the VEV $w$ also
breaks the 3-3-1 gauge symmetry down to that of the standard
model, and provides the masses for the exotic quarks $U$ and $D$
as well as the new gauge bosons. The $u,u'$ as well as $v,v'$
break the standard model symmetry, and give the masses for the
ordinary quarks, charged leptons and gauge bosons. To keep a
consistency with the effective theory, we assume that $w$ is much
larger than those of $\phi$ and $\eta$. In the following we
consider the first case of the unbroken parity.

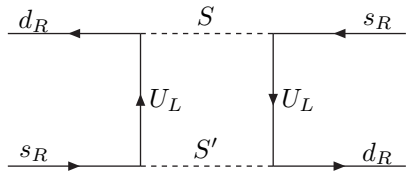
\begin{figure}[h]
\bc
\begin{picture}(200,100)(0,0)
\ArrowLine(50,25)(100,25)\ArrowLine(100,25)(100,75)\ArrowLine(100,75)(50,75)
\DashLine(100,25)(150,25){2}\DashLine(100,75)(150,75){2}
\ArrowLine(150,25)(200,25)\ArrowLine(150,75)(150,25)\ArrowLine(200,75)(150,75)
\Text(110,50)[]{$U_L$}\Text(160,50)[]{$U_L$} \Text(60,30)[]{$s_R$}
\Text(60,80)[]{$d_R$}\Text(190,80)[]{$s_R$} \Text(190,30)[]{$d_R$}
\Text(125,82)[]{$S$} \Text(125,32)[]{$S'$}
\end{picture}
\caption{\label{h1} A contribution to $K^0-\bar{K}^0$ mixing,
where $S,S'$ are respectively some combinations of the
singly-charged scalars, the vertices proportional to $h^d$ (or
$h'^d$) and appropriate mixing matrix elements.} \ec
\end{figure}

The exotic quarks get masses $m_U=f_3 w$ and $m_{D_{1,2}}=f w$,
where the $U$ and $D_{1,2}$ by themselves are the mass
eigenstates. The mass matrices for ordinary up-quarks and
down-quarks are, respectively, obtained as follows: \bea M_u &=&
\left(%
\begin{array}{ccc}
  -(h^u v+h'^u v') & -(h^u v+h'^u v') \om^2   & -(h^u v+h'^u v') \om   \\
   -(h^u v-h'^u v') &-(h^u v-h'^u v') \om  & -(h^u v-h'^u v') \om^2   \\
  h^u_3 u & h^u_3 u & h^u_3 u \\
\end{array}%
\right),\\
M_d &=&
\left(%
\begin{array}{ccc}
  h^d u+h'^d u' & (h^d u+h'^d u') \om^2   & (h^d u+h'^d u') \om   \\
   h^d u-h'^d u' &(h^d u-h'^d u') \om  & (h^d u-h'^d u') \om^2   \\
  h^d_3 v & h^d_3 v & h^d_3 v \\
\end{array}%
\right).\eea Let us define \bea A=
\fr{1}{\sqrt{3}}\left(%
\begin{array}{ccc}
  1 & 1 & 1 \\
  \om & \om^2 & 1 \\
  \om^2 & \om & 1 \\
\end{array}%
\right).\eea We have then \bea M_u A&=&
\left(%
\begin{array}{ccc}
  -\sqrt{3}(h^u v+h'^u v') & 0 & 0 \\
  0 & -\sqrt{3}(h^u v-h'^u v') & 0 \\
  0 & 0 & \sqrt{3}h^u_3 u \\
\end{array}%
\right)=\left(%
\begin{array}{ccc}
  m_u & 0 & 0 \\
  0 & m_c & 0 \\
  0 & 0 & m_t \\
\end{array}%
\right), \crn M_d A&=&
\left(%
\begin{array}{ccc}
  \sqrt{3}(h^d u+h'^d u') & 0 & 0 \\
  0 & \sqrt{3}(h^d u-h'^d u') & 0 \\
  0 & 0 & \sqrt{3}h^d_3 v \\
\end{array}%
\right)
=\left(%
\begin{array}{ccc}
  m_d & 0 & 0 \\
  0 & m_s & 0 \\
  0 & 0 & m_b \\
\end{array}%
\right).\eea In similarity to the charged leptons, the $u$ and $c$
quarks are also separated by the $\phi'$ scalar. We see also that
the introduction of $\eta'$ is necessary to provide the different
masses for $d$ and $s$ quarks. The unitary matrices, which couple
the left-handed up- and down-quarks to those in the mass bases,
are $U^u_L=1$ and $U^d_L=1$, respectively. Therefore we get the
quark mixing matrix \bea U_\mathrm{CKM}=U^{d\dagger}_L
U^u_L=1.\label{a41}\eea This is also an important result of our
paper since the experimental quark mixing matrix is close to the
unit matrix. In this case, the flavor changing neutral current
(FCNC) can arise from one-loop processes with the exchange of
heavy exotic quarks: see, for example, a contribution to the
$K^0-\bar{K}^0$ mixing due to the box diagram in Fig. \ref{h1}.
The amplitude after integrating out the heavy particles is
proportional to $[(h^d)^4/(16\pi^2 m^2_U)](\bar{d}_R \ga^\mu
s_R)(\bar{d}_R \ga_\mu s_R)$, which is strongly suppressed by the
loop factor and the exotic quark mass. The deviation of the CKM
matrix from the identity can be given by the FCNC effects with the
left-handed quarks, but such deviations are highly suppressed by
the mass of the extra quarks also.

If the lepton parity is spontaneously broken, i.e. $\langle
\eta^0_3\rangle$, $\langle \eta'^0_3\rangle$,
$\langle\chi^0_1\rangle$ $\neq 0$, then there exist the following
effects: (i) the mixings between ordinary quarks and exotic quarks
(namely, $u_{1,2,3}$ mix with $U$ and $d_{1,2,3}$ with $D_{1,2}$)
which can lead to FCNC processes at the tree level; (ii) the
result (\ref{a41}) is no longer correct, and the CKM is not
unitary. A small mixing among the ordinary quarks may exist due to
this violation. Let us recall that in the ordinary 3-3-1 model
without $S_4$, the Yukawa interactions like (\ref{yv}) might
additionally contain $\mathcal{L}$ explicitly-violating terms
\cite{clong}, which can be also the source contributing into the
ones similar to (i) and (ii). Such kinds of the mixings in the
3-3-1 model have been studied in a number of papers \cite{qmixng},
so we will not discuss it further. We remark that the mixings will
be very small since the parity breaking VEVs are strongly
suppressed by the same reason like $v_{\sigma,s}$ in (\ref{sI})
due to violating potentials. Anyway, the solution corresponding to
the residual symmetry $P_l$ as in the first case should be more
natural.

\section{\label{vev}Vacuum Alignment}

We can separate the general scalar potential into \bea
V_{\mathrm{total}}=V_{\mathrm{tri}}+V_{\mathrm{sext}}+V_{\mathrm{tri-sext}}+\overline{V},\eea
where $V_{\mathrm{tri}}$ and $V_{\mathrm{sext}}$ respectively
consist of the $\mathrm{SU}(3)_L$ scalar triplets and sextets,
whereas $V_{\mathrm{tri-sext}}$ contains the terms connecting the
two sectors. Moreover $V_{\mathrm{tri,sext,tri-sext}}$ conserve
$\mathcal{L}$--charge and $S_4$ symmetry, while $\overline{V}$
includes possible soft-terms explicitly violating these charges.
Here the soft-terms as we meant include the trilinear and quartic
ones as well. The reason for imposing $\bar{V}$ will be shown
below.

The details on the potentials are given as follows. We first
denote
$V(\textit{X}\rightarrow\textit{X}_1,\textit{Y}\rightarrow\textit{Y}_1,\cdots)
\equiv V(X,Y,\cdots)\!\!\!\mid_{X=X_1,Y=Y_1,\cdots}$ Notice also
that
$(\Tr\textit{A})(\Tr\textit{B})=\Tr(\textit{A}\Tr\textit{B})$.
$V_{\mathrm{tri}}$ is a sum of \bea
V(\chi)&=&\mu_{\chi}^2\chi^\+\chi
+\lambda^{\chi}({\chi}^\+\chi)^2,\\
V(\phi)&=&\mu_\phi^2(\phi^\+\phi)_{\underline{1}}
+\lambda_1^\phi(\phi^\+\phi)_{\underline{1}}(\phi^\+\phi)_{\underline{1}}
+\lambda_2^\phi(\phi^\+\phi)_{\underline{2}}(\phi^\+\phi)_{\underline{2}}\crn
&&+\lambda_3^\phi(\phi^\+\phi)_{\underline{3}}(\phi^\+\phi)_{\underline{3}}
+\lambda_4^\phi(\phi^\+\phi)_{\underline{3}'}(\phi^\+\phi)_{\underline{3}'},\\
V(\phi')&=&V(\phi\rightarrow \phi'),\hs V(\eta)=V(\phi\rightarrow
\eta),\hs V(\eta')=V(\phi\rightarrow \eta'),\\
V(\phi,\chi)&=&\lambda_1^{\phi\chi}({\phi}^\+\phi)_{\underline{1}}({\chi}^\+\chi)
+\lambda_2^{\phi\chi}({\phi}^\+\chi)(\chi^\+\phi),\\
V(\phi',\chi)&=&V(\phi\rightarrow \phi',\chi),\hs
V(\eta,\chi)=V(\phi\rightarrow \eta,\chi),\hs
V(\eta',\chi)=V(\phi\rightarrow \eta',\chi), \\
V(\phi,\eta)&=&{\lambda_1^{\phi\eta}}({\phi}^\+\phi)_{\underline{1}}({\eta}^\+\eta)_{\underline{1}}
+{\lambda_2^{\phi\eta}}({\phi}^\+\phi)_{\underline{2}}({\eta}^\+\eta)_{\underline{2}}
+{\lambda_3^{\phi\eta}}({\phi}^\+\phi)_{\underline{3}}({\eta}^\+\eta)_{\underline{3}}\crn
&&+{\lambda_4^{\phi\eta}}({\phi}^\+\phi)_{\underline{3}'}({\eta}^\+\eta)_{\underline{3}'}
+{\lambda_5^{\phi\eta}}({\eta}^\+\phi)_{\underline{1}}({\phi}^\+\eta)_{\underline{1}}
+{\lambda_6^{\phi\eta}}({\eta}^\+\phi)_{\underline{2}}({\phi}^\+\eta)_{\underline{2}}\crn
&&+{\lambda_7^{\phi\eta}}({\eta}^\+\phi)_{\underline{3}}({\phi}^\+\eta)_{\underline{3}}
+{\lambda_8^{\phi\eta}}({\eta}^\+\phi)_{\underline{3}'}({\phi}^\+\eta)_{\underline{3}'},\\
V(\phi',\eta')&=&V(\phi\rightarrow\phi',\eta\rightarrow\eta'),\\
V(\phi,\eta')&=&{\lambda_1^{\phi\eta'}}({\phi}^\+\phi)_{\underline{1}}({\eta'}^\+\eta')_{\underline{1}}
+{\lambda_2^{\phi\eta'}}({\phi}^\+\phi)_{\underline{2}}({\eta'}^\+\eta')_{\underline{2}}
+{\lambda_3^{\phi\eta'}}({\phi}^\+\phi)_{\underline{3}}({\eta'}^\+\eta')_{\underline{3}}\crn
&&+{\lambda_4^{\phi\eta'}}({\phi}^\+\phi)_{\underline{3}'}({\eta'}^\+\eta')_{\underline{3}'}
+{\lambda_5^{\phi\eta'}}({\eta'}^\+\phi)_{\underline{1}'}({\phi}^\+\eta')_{\underline{1}'}
+{\lambda_6^{\phi\eta'}}({\eta'}^\+\phi)_{\underline{2}}({\phi}^\+\eta')_{\underline{2}}\crn
&&+{\lambda_7^{\phi\eta'}}({\eta'}^\+\phi)_{\underline{3}}({\phi}^\+\eta')_{\underline{3}}
+{\lambda_8^{\phi\eta'}}({\eta'}^\+\phi)_{\underline{3}'}({\phi}^\+\eta')_{\underline{3}'},\\
V(\phi',\eta)&=&V(\phi\rightarrow\phi',\eta'\rightarrow\eta),\\
V(\phi,\phi')&=&V(\phi\rightarrow \phi,\eta'\rightarrow
\phi')+\left[{\lambda_9^{\phi\phi'}}({\phi}^\+\phi)_{\underline{2}}({\phi}^\+\phi')_{\underline{2}}
+{\lambda_{10}^{\phi\phi'}}({\phi}^\+\phi)_{\underline{3}}({\phi}^\+\phi')_{\underline{3}}
+{\lambda_{11}^{\phi\phi'}}({\phi}^\+\phi)_{\underline{3}'}({\phi}^\+\phi')_{\underline{3}'}\right.
\crn &&+\left.
{\lambda_{12}^{\phi\phi'}}({\phi'}^\+\phi')_{\underline{2}}({\phi'}^\+\phi)_{\underline{2}}
+{\lambda_{13}^{\phi\phi'}}({\phi'}^\+\phi')_{\underline{3}}({\phi'}^\+\phi)_{\underline{3}}
+{\lambda_{14}^{\phi\phi'}}({\phi'}^\+\phi')_{\underline{3}'}({\phi'}^\+\phi)_{\underline{3}'}+h.c.\right],\crn
V(\eta,\eta')&=&V(\phi\rightarrow\eta,\phi'\rightarrow\eta'),\\
V_{\chi\phi\phi'\eta\eta'}&=&{\mu_1}\chi\phi\eta+{\mu'_1}\chi\phi'\eta'
\crn
&&+{\lambda^1_1}(\phi^+\phi')_{\underline{1}'}(\eta^+\eta')_{\underline{1}'}
+{\lambda^1_2}(\phi^\+\phi')_{\underline{2}}(\eta^\+\eta')_{\underline{2}}
+{\lambda^1_3}(\phi^\+\phi')_{\underline{3}}(\eta^\+\eta')_{\underline{3}}
+{\lambda^1_{4}}(\phi^\+\phi')_{\underline{3}'}(\eta^\+\eta')_{\underline{3}'}
\crn &&
+{\lambda}^2_1(\phi^\+\eta')_{\underline{1}'}(\eta^\+\phi')_{\underline{1}'}
+{\lambda}^2_2(\phi^\+\eta')_{\underline{2}}(\eta^\+\phi')_{\underline{2}}+
{\lambda}^2_3(\phi^\+\eta')_{\underline{3}}(\eta^\+\phi')_{\underline{3}}
+{\lambda}^2_4(\phi^\+\eta')_{\underline{3}'}(\eta^\+\phi')_{\underline{3}'}
\crn &&
+{\lambda^3_1}(\phi^+\phi')_{\underline{1}'}(\eta'^+\eta)_{\underline{1}'}
+{\lambda^3_2}(\phi^\+\phi')_{\underline{2}}(\eta'^\+\eta)_{\underline{2}}
+{\lambda^3_3}(\phi^\+\phi')_{\underline{3}}(\eta'^\+\eta)_{\underline{3}}
+{\lambda^3_{4}}(\phi^\+\phi')_{\underline{3}'}(\eta'^\+\eta)_{\underline{3}'}
\crn &&
+{\lambda}^4_1(\phi^\+\eta)_{\underline{1}}(\eta'^\+\phi')_{\underline{1}}
+{\lambda}^4_2(\phi^\+\eta)_{\underline{2}}(\eta'^\+\phi')_{\underline{2}}+
{\lambda}^4_3(\phi^\+\eta)_{\underline{3}}(\eta'^\+\phi')_{\underline{3}}
+{\lambda}^4_4(\phi^\+\eta)_{\underline{3}'}(\eta'^\+\phi')_{\underline{3}'}
\crn &&
+{\lambda^5_1}(\phi^+\eta)_{\underline{1}}(\phi'^+\eta')_{\underline{1}}
+{\lambda^5_2}(\phi^+\eta)_{\underline{2}}(\phi'^+\eta')_{\underline{2}}
+{\lambda^5_3}(\phi^+\eta)_{\underline{3}}(\phi'^+\eta')_{\underline{3}}
+{\lambda^5_{4}}(\phi^+\eta)_{\underline{3}'}(\phi'^+\eta')_{\underline{3}'}
\crn &&
+{\lambda}^6_1(\phi^\+\eta')_{\underline{1}'}(\phi'^\+\eta)_{\underline{1}'}
+{\lambda}^6_2(\phi^\+\eta')_{\underline{2}}(\phi'^\+\eta)_{\underline{2}}+
{\lambda}^6_3(\phi^\+\eta')_{\underline{3}}(\phi'^\+\eta)_{\underline{3}}
+{\lambda}^6_4(\phi^\+\eta')_{\underline{3}'}(\phi'^\+\eta)_{\underline{3}'}\crn
&&+h.c. \eea $V_{\mathrm{sext}}$ is a sum of
 \bea
 V(\sigma)&=&\Tr[V(\chi\rightarrow\sigma)+\lambda'^{\sigma}(\sigma^\+
\sigma)\Tr({\sigma}^\+\sigma)],\\
V(s)&=&\Tr\{V(\phi\rightarrow{s})+{\lambda'}_1^{s}(s^\+s)_{\underline{1}}\Tr(s^\+s)_{\underline{1}}
+{\lambda'}_2^{s}(s^\+s)_{\underline{2}}\Tr(s^\+s)_{\underline{2}}\crn
&&+{\lambda'}_3^{s}(s^\+s)_{\underline{3}}\Tr(s^\+s)_{\underline{3}}
+{\lambda'}_4^{s}(s^\+s)_{\underline{3}'}\Tr(s^\+s)_{\underline{3}'}\},\\
V(s,\sigma)&=&\Tr\{V(\phi\rightarrow{s},\chi\rightarrow\sigma)
+{\lambda_1'}^{s\sigma}(s^\+s)_{\underline{1}}\Tr(\sigma^\+\sigma)
+{\lambda_2'}^{s\sigma}(s^\+\sigma)\Tr({\sigma^\+}s)\crn &&
+[{\lambda_3'}^{s\sigma}(s^\+\sigma)\Tr(s^\+\sigma)
+{\lambda_4'}^{s\sigma}({\sigma^\+}s)\Tr({s^\+}s)_{\underline{3}}
+h.c.]\}\eea $V_{\mathrm{tri-sext}}$ is a sum of \bea
V(\sigma,\chi)&=&\lambda_1^{\sigma\chi}({\chi}^\+\chi)\Tr({\sigma}^\+\sigma)
+\lambda_2^{\sigma\chi}{({\chi}^\+\sigma^\dagger)({\sigma}\chi)}
+(\mu_2\chi^T\sigma\chi+h.c.),\\
V(s,\chi)&=&\Tr[V(\phi\rightarrow{s}^\+,\chi\rightarrow\chi)],\hs
V(\phi,\sigma)=\Tr[V(\phi\rightarrow\phi,\chi\rightarrow\sigma^\+)],\\
V(\phi,s)&=&\Tr[V(\phi\rightarrow\phi,\eta\rightarrow{s}^\+)],\hs
V(\phi',\sigma)=\Tr[V(\phi'\rightarrow\phi',\chi\rightarrow\sigma^\+)],\\
V(\phi',s)&=&\Tr[V(\phi'\rightarrow\phi',\eta\rightarrow{s}^\dagger)],\hs
V(\eta,\sigma)=\Tr[V(\eta\rightarrow\eta,\chi\rightarrow\sigma^\dagger)],\\
V(\eta,s)&=&\Tr[V(\phi\rightarrow\eta,\eta\rightarrow{s}^\+)],\hs
V(\eta',\sigma)=\Tr[V(\eta'\rightarrow\eta',\chi\rightarrow\sigma^\+)],\\
V(\eta',s)&=&\Tr[V(\phi'\rightarrow\eta',\eta\rightarrow s^\+)],\\
V_{s\sigma\chi \phi\phi' \eta \eta'}&=&\chi^\+ \sigma^\+
(\lambda_1 \phi\eta+ \lambda_2 \phi'\eta')_{\underline{1}}+\chi^\+
s^\+ (\la_3 \phi\eta + \la_4 \phi' \eta' +\la_5 \phi \eta' +\la_6
\phi'\eta)_{\underline{3}}\crn &&+\Tr(s^\+s)_{\underline{2}}(\la_7
\phi^\+\phi'+\la_8 \eta^\+\eta')_{\underline{2}}
+\Tr(s^\+s)_{\underline{3}}(\la_9 \phi^\+\phi'+\la_{10}
\eta^\+\eta')_{\underline{3}}\crn
&&+\Tr(s^\+s)_{\underline{3}'}(\la_{11} \phi^\+\phi'+\la_{12}
\eta^\+\eta')_{\underline{3}'} +\Tr(\sigma^\+
s)(\la_{13}\phi^\+\phi +\la_{14}\phi'^\+\phi'\crn && +
\la_{15}\phi^\+\phi' + \la_{16}\phi'^\+\phi +\la_{17}\eta^\+\eta+
\la_{18}\eta'^\+\eta'+\la_{19}\eta^\+\eta' +
\la_{20}\eta'^\+\eta)_{\underline{3}} \crn &&+\phi^\+\sigma^\+s
(\la_{21} \phi+\la_{22}\phi')+\phi'^\+\sigma^\+s (\la_{23}
\phi+\la_{24}\phi')+\eta^\+\sigma^\+s (\la_{25}
\eta+\la_{26}\eta')\crn && +\eta'^\+\sigma^\+s (\la_{27}
\eta+\la_{28}\eta')+\la_{29}(\phi^\+ s^\+)_{\underline{2}} (s
\phi')_{\underline{2}}+\la_{30}(\phi^\+ s^\+)_{\underline{3}} (s
\phi')_{\underline{3}}\crn &&+\la_{31}(\phi^\+
s^\+)_{\underline{3}'} (s \phi')_{\underline{3}'}
+\la_{32}(\eta^\+ s^\+)_{\underline{2}} (s
\eta')_{\underline{2}}+\la_{33}(\eta^\+ s^\+)_{\underline{3}} (s
\eta')_{\underline{3}}\crn &&+\la_{34}(\eta^\+
s^\+)_{\underline{3}'} (s \eta')_{\underline{3}'}+h.c.\eea And,
the $\overline{V}$ up to quartic interactions is given by \bea
\bar{V}&=& (\bar{\mu}_1\eta \eta+\bar{\mu}'_1\eta'
\eta')_{\underline{1}}\sigma+ (\bar{\mu}_2\eta
\eta+\bar{\mu}'_2\eta'\eta'+\bar{\mu}''_2\eta\eta'+\bar{\mu}_3\chi\eta)_{\underline{3}}s\crn
&& + \eta^\+\sigma^\+(\bar{\la}_1\phi\chi +
\bar{\la}_2\phi\eta+\bar{\la}_3 \phi'\eta'+\bar{\la}_4\phi'\eta
+\bar{\la}_5\phi \eta')_{\underline{3}}+
\eta'^\+\sigma^\+(\bar{\la}_6\phi'\chi +
\bar{\la}_7\phi\eta+\bar{\la}_8 \phi'\eta'\crn
&&+\bar{\la}_9\phi'\eta +\bar{\la}_{10}\phi
\eta')_{\underline{3}'}+\bar{\la}_{11}\phi^\+\sigma^\+(\phi\phi')_{\underline{3}}
+\bar{\la}_{12}\phi'^\+\sigma^\+(\phi\phi')_{\underline{3}'}+\bar{\la}_{13}\chi^\+s^\+\phi\chi
\crn &&
+(\eta^\+s^\+)_{\underline{1}}(\bar{\la}_{14}\phi\eta+\bar{\la}_{15}\phi'\eta')_{\underline{1}}
+(\eta^\+s^\+)_{\underline{2}}(\bar{\la}_{16}\phi\eta+\bar{\la}_{17}\phi'\eta'
+\bar{\la}_{18}\phi'\eta+\bar{\la}_{19}\phi\eta')_{\underline{2}}
\crn &&+(\eta^\+s^\+)_{\underline{3}}(\bar{\la}_{20}\phi\eta
+\bar{\la}_{21}\phi'\eta'+\bar{\la}_{22}\phi'\eta+\bar{\la}_{23}\phi\eta')_{\underline{3}}
 +(\eta^\+s^\+)_{\underline{3}'}(\bar{\la}_{24}\phi\eta
+\bar{\la}_{25}\phi'\eta'\crn &&
+\bar{\la}_{26}\phi'\eta+\bar{\la}_{27}\phi\eta')_{\underline{3}'}
+(\eta'^\+s^\+)_{\underline{1}'}(\bar{\la}_{28}\phi'\eta
+\bar{\la}_{29}\phi\eta')_{\underline{1}'}+(\eta'^\+s^\+)_{\underline{2}}(\bar{\la}_{30}\phi\eta
+\bar{\la}_{31}\phi'\eta'\crn &&+\bar{\la}_{32}\phi'\eta
+\bar{\la}_{33}\phi\eta')_{\underline{2}}
+(\eta'^\+s^\+)_{\underline{3}}(\bar{\la}_{34}\phi\eta
+\bar{\la}_{35}\phi'\eta'+\bar{\la}_{36}\phi'\eta+\bar{\la}_{37}\phi\eta')_{\underline{3}}
\crn && +(\eta'^\+s^\+)_{\underline{3}'}(\bar{\la}_{38}\phi\eta
+\bar{\la}_{39}\phi'\eta'
+\bar{\la}_{40}\phi'\eta+\bar{\la}_{41}\phi\eta')_{\underline{3}'}+\bar{\la}_{42}(\phi'^\+s^\+)_{\underline{1}'}
(\phi\phi')_{\underline{1}'}\crn
&&+\bar{\la}_{43}(\phi'^\+s^\+)_{\underline{2}}
(\phi\phi')_{\underline{2}}+\bar{\la}_{44}(\phi'^\+s^\+)_{\underline{3}}
(\phi\phi')_{\underline{3}}+\bar{\la}_{45}(\phi'^\+s^\+)_{\underline{3}'}
(\phi\phi')_{\underline{3}'}+\bar{\la}_{46}(\phi^\+s^\+)_{\underline{2}}
(\phi\phi')_{\underline{2}}\crn
&&+\bar{\la}_{47}(\phi^\+s^\+)_{\underline{3}}
(\phi\phi')_{\underline{3}}+\bar{\la}_{48}(\phi^\+s^\+)_{\underline{3}'}
(\phi\phi')_{\underline{3}'}+[\bar{\la}_{49}\Tr(s^\+s)+\bar{\la}_{50}\Tr(s^\+\sigma)+\bar{\la}_{51}\Tr(\sigma^\+s)
\crn && +\bar{\la}_{52}\eta^\+\chi
+\bar{\la}_{53}\eta^\+\eta+\bar{\la}_{54}\eta'^\+\eta'+\bar{\la}_{55}\eta^\+\eta'+\bar{\la}_{56}\eta'^\+\eta
 +\bar{\la}_{57}\phi^\+\phi+\bar{\la}_{58}\phi'^\+\phi'
+\bar{\la}_{59}\phi^\+\phi'\crn
&&+\bar{\la}_{60}\phi'^\+\phi]_{\underline{3}}\eta^\+\chi
 +[\bar{\la}_{61}\Tr(s^\+s)+\bar{\la}_{62}\eta'^\+\chi
+\bar{\la}_{63}\eta^\+\eta+\bar{\la}_{64}\eta'^\+\eta'+\bar{\la}_{65}\eta^\+\eta'+\bar{\la}_{66}\eta'^\+\eta
\crn &&
+\bar{\la}_{67}\phi^\+\phi+\bar{\la}_{68}\phi'^\+\phi'+\bar{\la}_{69}\phi^\+\phi'
+\bar{\la}_{70}\phi'^\+\phi]_{\underline{3}'}\eta'^\+\chi
+\bar{\la}_{71}(\eta^\+\phi)_{\underline{3}}(\phi^\+\chi)+\bar{\la}_{72}(\eta^\+\phi')_{\underline{3}'}
(\phi'^\+\chi)\crn &&
+\bar{\la}_{73}(\eta^\+\phi)_{\underline{3}'}(\phi'^\+\chi)+\bar{\la}_{74}(\eta^\+\phi')_{\underline{3}}(\phi^\+\chi)+
\bar{\la}_{75}(\eta'^\+\phi)_{\underline{3}}(\phi^\+\chi)+\bar{\la}_{76}(\eta'^\+\phi')_{\underline{3}'}
(\phi'^\+\chi) \crn &&
+\bar{\la}_{77}(\eta'^\+\phi)_{\underline{3}'}(\phi'^\+\chi)+\bar{\la}_{78}(\eta'^\+\phi')_{\underline{3}}(\phi^\+\chi)
+\bar{\la}_{79}(\eta^\+s^\+)_{\underline{3}}s\chi
+\bar{\la}_{80}(\eta'^\+s^\+)_{\underline{3}}s\chi+\bar{\la}_{81}\eta^\+s^\+\sigma\chi\crn
&&+\bar{\la}_{82}\eta^\+\sigma^\+s\chi+h.c.,\label{vi}\eea where
all the terms in this potential violate the $\mathcal{L}$-charge,
but conserving $S_4$. Yet we have not pointed out, but there must
additionally exist the terms in $\bar{V}$ explicitly violating the
only $S_4$ symmetry or both the $S_4$ and $\mathcal{L}$-charge
too. In the following, most of them will be omitted, only the
terms of the kind of interest are provided.

There are the several scalar sectors corresponding to the expected
VEV directions: $(1,0,0)$ for $s$ and $(1,1,1)$ for $\phi,\
\phi',\ \eta,\ \eta'$, as written out before. However if these
sectors are strongly coupled through the potential
$V_{\mathrm{tri-sext}}\neq 0$, such vacuum misalignment cannot be
given from the potential minimization. To overcome the difficulty,
as in the literature we might include the extradimensions or
supersymmetry, or using additional discrete symmetries. However,
in this paper we will provide an alternative explanation,
following the works in Refs.\cite{A4} of Ma and/nor collaborations
in 2001, 2004, and 2010. We thus suppose that $\sigma$ and $s$ are
all very heavy (see also \cite{matd}) with masses $\mu_\sigma$ and
$\mu_s$ respectively, so that all of them (as given in
$V_{\mathrm{tri-sext}}$) are integrated away. They therefore have
the only interactions among themselves as given in
$V_{\mathrm{sext}}$. They do not appear as physical particles at
or below the TeV scale. Only their imprint at the low energy is a
resulting effective potential, which consists of only the fields
$\phi$, $\phi'$, $\eta$, $\eta'$ and $\chi$, up to the fourth
orders having the same form as $V_{\mathrm{tri}}$.

Consider the potential $V_{\mathrm{tri}}$. The flavons $\phi\ ,\
\phi',\ \eta,\ \eta'$ with their VEVs aligned in the same
direction $(1,1,1)$ are a automatical solution from the
minimization conditions of $V_{\mathrm{tri}}$. To see this
obviously, in the system of minimization equations let us put
$v_1=v_2=v_3=v$, $v'_1=v'_2=v'_3=v'$, $u_1=u_2=u_3=u$, and
$u'_1=u'_2=u'_3=u'$, which reduces to
\bea(\mu_\phi^2+\lambda_1^{\phi\chi}v_{\chi}^2)v+(3\lambda_1^{\phi\eta}
+4\lambda_3^{\phi\eta})u^2v+(3\lambda_1^{\phi\eta'}+4\lambda_3^{\phi\eta'})u'^2v
+(6\lambda_1^{\phi}+8\lambda_3^{\phi})v^3\crn
+(3\lambda_1^{\phi\phi'}+4\lambda_3^{\phi\phi'}+3{\lambda}_5^{\phi\phi'}
+4{\lambda}_8^{\phi\phi'})vv'^2+(3\lambda_1^1+4\lambda_4^1+3{\lambda}_1^3+4{\lambda}_4^3)u
u'v'=0,\\
(\mu_{\phi'}^2+\lambda_1^{\phi'\chi}v_{\chi}^2)v'+(3\lambda_1^{\phi'\eta}
+4\lambda_3^{\phi'\eta})u^2v'+(3{\lambda}_1^{\phi'\eta'}+4{\lambda}_3^{\phi'\eta'})u'^2v'
+(6\lambda_1^{\phi'}+8\lambda_3^{\phi'})v'^3\crn+(3{\lambda}_1^{\phi\phi'}
+4{\lambda}_3^{\phi\phi'}+3{\lambda}_5^{\phi\phi'}+4{\lambda}_8^{\phi\phi'})v^2v'
+(3\lambda_1^1+4\lambda_4^1+3\lambda_1^3+4\lambda_4^3)uu'v=0,\\
(\mu_\eta^2+\lambda_1^{\chi\eta}v_{\chi}^2)u+(3\lambda_1^{\phi\eta}
+4\lambda_3^{\phi\eta})v^2u+(3\lambda_1^{\phi'\eta}+4\lambda_3^{\phi'\eta})v'^2u
+(6\lambda_1^{\eta}+8\lambda_3^{\eta})u^3\crn
+(3\lambda_1^{\eta\eta'}+4\lambda_3^{\eta\eta'}+3{\lambda}_5^{\eta\eta'}+4{\lambda}_8^{\eta\eta'})u'^2u
+(3\lambda_1^1+4\lambda_4^1)u'v'v=0,\\
(\mu_{\eta'}^2+\lambda_1^{\eta'\chi}v_{\chi}^2)u'+(3\lambda_1^{\phi\eta'}
+4\lambda_3^{\phi\eta'})u'v^2+(3\lambda_1^{\phi'\eta'}+4\lambda_3^{\phi'\eta'})u'v'^2
+(6\lambda_1^{\eta'}+8\lambda_3^{\eta'})u'^3\crn+(3\lambda_1^{\eta\eta'}
+4\lambda_3^{\eta\eta'}+3{\lambda}_5^{\eta\eta'}+4{\lambda}_8^{\eta\eta'})u^2u'
+(3\lambda_1^3+4\lambda_4^3)uvv'=0.\eea This system always give
the solution ($u,v,u',v'$) as expected, even though it is
complicated. It is also noted that the aligned $(1,1,1)$ as given
is only one solution. The other directions such as $(1,0,0)$ are
also the solution of the potential minimization. We have thus
imposed the first case to have the desirable results.

Now we consider the potential $V^{s\sigma}$ concerning the
antisextets. To obtain the desirable solution
$\langle\sigma\rangle \neq 0$, $\langle s_1\rangle \neq 0$, and
$\langle s_2\rangle = \langle s_3\rangle =0$, the
$\mathcal{L}$-charge as well as the $S_4$ symmetry must be broken
as spoken of around (\ref{vi}). Assume the following choice of
soft scalar trilinear and quartic terms as given in the general
potential expression $\bar{V}$ works in $V^{s\sigma}$:\bea
V^{s\sigma}&=&V_{\mathrm{sext}}+ [\bar{\mu}_1(\eta
\eta)_{\underline{1}}\sigma+ \bar{\mu}_2(\eta
\eta)_{\underline{1}}s_1 + \bar{\la}_1
\eta^\+\sigma^\+(\phi\eta)_{\underline{3}}+
\bar{\la}_{2}\eta^\+s_1^\+(\phi\eta)_{\underline{3}} +h.c.]\eea To
understand this, note first that in order for $\sigma$ or
$s_{1,2,3}$ to have a VEV, $\mathcal{L}$ must be broken and that
can only be achieved through the terms of $\bar{V}$. However, as
in one of the works of Ma cited above, we can introduce a protect
symmetry $Z_2$ so that the $s_2$ and $s_3$ are only connected to
the terms in the potentials or the Yukawa couplings, which always
preserve the symmetry $\psi_{2,3}\rightarrow -\psi_{2,3}$, where
$\psi$ is any $S_4$ triplet appearing in the text such as
$s,\phi,\psi_L$ and so on. Hence they always appear together and
protect each other from getting a VEV if neither has one to begin
with.

From $V^{s\sigma}$, the unique solution to the minimization
conditions is $\langle s_2\rangle =\langle s_3\rangle =0$ and
nonzero but very small values of $\lambda_{\sigma,s}$ and
$v_{\sigma,s}$ as induced in $\langle s_1\rangle $ and $\langle
\sigma\rangle$ of Eqs. (\ref{s1},\ref{sim}) being the root of the
$\pa V^{s\sigma}_{\mathrm{min}}/\pa \langle s_1\rangle ^*=0$ and
$\pa V^{s\sigma}_{\mathrm{min}}/\pa \langle \sigma\rangle ^*=0$
(with $V^{s\sigma}_{\mathrm{min}}$ the minimum of $V^{s\sigma}$).
First, the equations $\pa V^{s\sigma}_{\mathrm{min}}/\pa
\La_\sigma^*=0$ and $\pa V^{s\sigma}_{\mathrm{min}}/\pa \La_s
^*=0$ imply that $\La_\sigma$ and $\La_s$ are in the scale of the
antisextets' masses $\mu_\sigma$ and $\mu_s$ \cite{dlsh}. Let us
denote a characteristic scale $M$ so that
$\La_\sigma,\La_s,\mu_\sigma,\mu_s\sim M$. The remaining equations
$\pa V^{s\sigma}_{\mathrm{min}}/\pa \lambda_{\sigma,s}^*=0$ and
$\pa V^{s\sigma}_{\mathrm{min}}/\pa v_{\sigma,s}^*=0$ provide the
small VEVs induced by the standard model electroweak scale $u\sim
v$:\bea \la_\sigma &\sim& \bar{\mu}_1\fr{v^2}{M^2},\hs
\la_s \sim \bar{\mu}_2\fr{v^2}{M^2},\label{sII}\\
v_\sigma &\sim& \bar{\la}_1v \fr{v^2}{M^2},\hs v_s \sim
\bar{\la}_2v\fr{v^2}{M^2}.\label{sI}\eea

The parameters $\bar{\mu}_{1,2}$ and $\bar{\la}_{1,2}v$ (which
have the dimension of mass) may be naturally small in comparison
with $v$, because its absence enhances the symmetry of $V^{\sigma
s}$. We remark that the VEVs of the type II seesaw mechanism
$\la_\sigma$, $\la_s$ work because from (\ref{sII}) the
spontaneous breaking of electroweak symmetry is already
accomplished by $v$, the $\la_\sigma$, $\la_s$ may be small, as
long as $M$ is large. On the other hand, $v_\sigma$ and $v_s$ are
the VEVs of the type I seesaw mechanism which are also small for
the same reason; therefore, in this model the seesaw scale $M$ may
be much lower than that of the unusual type I seesaw. These are
also the important results of our paper.

Along the model, as mentioned the new particles are: $N_R$ getting
masses in $\La_{\sigma,s}$ scale, $U$ and $D$ with masses
proportional to $w$, and $Z'$, $X$, $Y$ having masses as
combinations of $w$ and $\La_{\sigma,s}$, where $w$ and
$\La_{\sigma,s}$ are the scales of 3-3-1 gauge symmetry breaking
down to the standard model \cite{331m,331r}. If the antisextets
$\sigma$, $s$ are heaviest, i.e. $\La^2_{\sigma,s}\gg w^2$, the
new gauge bosons and $N_R$ will have large masses ranging in this
scale accordingly, however $U$ and $D$ can gain masses much
smaller than (for example, in some hundreds of GeV). In the case
of $w\sim \La_{\sigma,s}$, the masses of $U$ and $D$ will be
picked up in the same order with those of the new gauge bosons and
$N_R$. By the way, the $\chi$ scalar may be also integrated out
like the antisextets. This will explain why the parity breaking
parameters $\langle \eta^0_3\rangle$, $\langle \eta'^0_3\rangle$,
$\langle\chi^0_1\rangle$ are small, in similarity to
$v_{\sigma,s}$. The mixings among the ordinary quarks and exotic
quarks and the tree-level FCNC as mentioned can be suppressed by
this mechanism.

There are a lot of $\mathrm{SU}(2)_L$ scalar doublets and triplets
in the model, under which they can lead to modifications for the
precision electroweak data (see \cite{longinami} for a detailed
analysis on this problem). The most serious one comes from
tree-level corrections for the $\rho$ parameter. In the effective
theory limit, the mass of $W$ boson and $\rho$ are evaluated by
\be m^2_W=\fr{g^2}{2}v^2_{\mathrm{w}},\hs \rho=\fr{m^2_W}{c^2_W
m^2_Z}=1-\fr{2(\la^2_{\sigma}+\la^2_s)}{v^2_{\mathrm{w}}},\ee
where $v^2_{\mathrm{w}}\simeq 3(v^2+v'^2+u^2+u'^2)=(174\
\mathrm{GeV})^2$ is a natural approximation due to
$v^2_{\sigma},v^2_{s}, \langle\chi^0_1\rangle^2\ll
v^2,v'^2,u^2,u'^2$, as given above. Because $\la_{\sigma,s}$ are
in eV order responsible for the neutrino masses, the $\rho$
parameter is absolutely close to $1$, which is in good agreement
with the data \cite{pdg}.

\section{\label{conclus}Conclusions}

As a result of anomaly cancelation, the 3-3-1 model accepts
discrepancy of one family of quarks from other two. We have
therefore searched for a symmetry group acting on both 2-family
and 3-family indices, the simplest of which is $S_4$---the
symmetry group of a cube as a flavor symmetry. Corresponding to
the lepton number, the new lepton charge $\mathcal{L}$ and its
residual symmetry---the lepton parity $P_l$ have been introduced
into the model.

If $P_l$ is conserved, the neutrino masses come from small VEVs of
first components of scalar antisextets, known as type II seesaw
contributions. If $P_l$ is broken there are additional
contributions from type I seesaw due to suppression of 3-3-1
symmetry breaking VEVs of just the antisextets. The tribimaximal
mixing arises as a result under $S_4$ and $\mathcal{L}$
symmetries. A deviation from this mixing can result from
$\mathcal{L}$ small violating terms or $S_4$ breaking soft-terms.
By imposing appropriate $\mathcal{L}$ and $S_4$ violating
potential, the VEV alignments have been obtained. Also, the
smallness of the seesaw contributions have been explained.

Quark mixing matrix is unity at the tree-level only if $P_l$ is
exact, not spontaneously broken. A breaking of the charge will
lead to mixings between exotic quarks and ordinary quarks. It can
also provide mixings among the ordinary quarks. In this case the
CKM is not unitary. There are contributions to flavor changing
neutral currents at the tree-level.

The model can provide interesting candidates for dark matter
without supersymmetry as stored in the antisextet flavons as well
as in the $\chi$ triplet if the lepton parity is conserved (see
also the notes as sketched in \cite{takahashi}), and the model's
phenomenology is very rich. They are worthy to be devoted to
further studies.

\section*{Acknowledgments}
We would like to thank Ryo Takahashi for his comments and showing
possible dark matter candidates existing in our model. We are
grateful to Yin Lin for communications and indicating us to some
papers in \cite{A4}. This work was supported in part by the
National Foundation for Science and Technology Development
(NAFOSTED) of Vietnam under Grant No. 103.01.15.09.
\\[0.3cm]

\appendix

\section{\label{apa}$\emph{S}_4$ group and Clebsch-Gordan coefficients}

$S_4$ is the permutation group of four objects, which is also the
symmetry group of a cube. It has 24 elements divided into 5
conjugacy classes, with \underline{1}, \underline{1}$'$,
\underline{2}, \underline{3}, and \underline{3}$'$ as its 5
irreducible representations. Any element of $S_4$ can be formed by
multiplication of the generators $S$ and $T$ obeying the relations
$S^4 = T^3 = 1,\ ST^2S = T.$ Without loss of generality, we could
choose $S = (1234),\ T = (123)$ where the cycle (1234) denotes the
permutation $(1, 2, 3, 4) \rightarrow (2, 3, 4, 1)$, and (123)
means $(1, 2, 3, 4) \rightarrow (2, 3, 1, 4)$. The conjugacy
classes generated from $S$ and $T$ are\bea C_1 &:& 1 \crn C_2 &:&
(12)(34)=TS^2T^2,\ (13)(24)=S^2,\ (14)(23)=T^2S^2T\crn C_3 &:&
(123)=T,\ (132)=T^2,\ (124)=T^2S^2,\ (142)=S^2T,\crn &&
(134)=S^2TS^2,\ (143)=STS,\ (234)=S^2T^2,\ (243)=TS^2\crn C_4 &:&
(1234)=S,\ (1243)=T^2ST,\ (1324)=ST,\crn && (1342)=TS,\
(1423)=TST^2,\ (1432)=S^3\crn C_5 &:& (12)=STS^2,\ (13)=TSTS^2,\
(14)=ST^2,\crn && (23)=S^2TS,\ (24)=TST,\ (34)=T^2S\nn \eea

The character table of $S_4$ is given as follows
\begin{center}
\begin{tabular}{|c|c|c|c|c|c|c|c|}
\hline Class & $n$ & $h$ & $\chi_{\underline{1}}$ &
$\chi_{\underline{1}'}$ & $\chi_{\underline{2}}$ &
$\chi_{\underline{3}}$ & $\chi_{\underline{3}'}$
\\
\hline
$C_1$ & 1 & 1 & 1 & 1 & 2 & 3 & 3 \\
$C_2$ & 3 & 2 & 1 & 1 & 2 & --1 & --1 \\
$C_3$ & 8 & 3 & 1 & 1 & --1 & 0 & 0 \\
$C_4$ & 6 & 4 & 1 & --1 & 0 & --1 & 1 \\
$C_5$ & 6 & 2 & 1 & --1 & 0 & 1 & --1 \\
\hline
\end{tabular}
\end{center}
where $n$ is the order of class and $h$ the order of elements
within each class. Let us note that $C_{1,2,3}$ are even
permutations, while $C_{4,5}$ are odd permutations. The two
three-dimensional representations differ only in the signs of
their $C_4$ and $C_5$ matrices. Similarly, the two one-dimensional
representations behave the same.

We will work in basis where $\underline{3},\underline{3}'$ are
real representations whereas $\underline{2}$ is complex. One
possible choice of generators is given as follows \bea
\underline{1}&:& S=1,\hs T=1 \crn \underline{1}'
&:& S=-1,\hs T=1 \crn \underline{2} &:& S=\left(%
\begin{array}{cc}
  0 & 1 \\
  1 & 0 \\
\end{array}%
\right),\hs T=\left(%
\begin{array}{cc}
  \om & 0 \\
  0 & \om^2 \\
\end{array}%
\right)\crn \underline{3}&:& S=\left(%
\begin{array}{ccc}
  -1 & 0 & 0 \\
  0 & 0 & -1 \\
  0 & 1 & 0 \\
\end{array}%
\right),\hs T=\left(%
\begin{array}{ccc}
  0 & 0 & 1 \\
  1 & 0 & 0 \\
  0 & 1 & 0 \\
\end{array}%
\right)\crn
\underline{3}'&:& S=-\left(%
\begin{array}{ccc}
  -1 & 0 & 0 \\
  0 & 0 & -1 \\
  0 & 1 & 0 \\
\end{array}%
\right),\hs T=\left(%
\begin{array}{ccc}
  0 & 0 & 1 \\
  1 & 0 & 0 \\
  0 & 1 & 0 \\
\end{array}%
\right)\eea where $\om=e^{2\pi i/3}=-1/2+i\sqrt{3}/2$ is the cube
root of unity. Using them we calculate the Clebsch-Gordan
coefficients for all the tensor products as given below.

First, let us put $\underline{3}(1,2,3)$ which means some
$\underline{3}$ multiplet such as $x=(x_1,x_2,x_3)\sim
\underline{3}$ or $y=(y_1,y_2,y_3)\sim \underline{3}$ or so on,
and similarly for the other representations. Moreover, the
numbered multiplets such as $(...,ij,...)$ mean $(...,x_i
y_j,...)$ where $x_i$ and $y_j$ are the multiplet components of
different representations $x$ and $y$, respectively. In the
following the components of representations in l.h.s will be
omitted and should be understood, but they always exist in order
in the components of decompositions in r.h.s: \bea
\underline{1}\otimes\underline{1}&=&\underline{1}(11),\hs
\underline{1}'\otimes \underline{1}'=\underline{1}(11),\hs
\underline{1}\otimes\underline{1}'=\underline{1}'(11),\\
\underline{1}\otimes \underline{2}&=&\underline{2}(11,12),\hs
\underline{1}'\otimes \underline{2}=\underline{2}(11,-12),\\
\underline{1}\otimes \underline{3}&=&\underline{3}(11,12,13),\hs
\underline{1}'\otimes \underline{3}=\underline{3}'(11,12,13),\\
\underline{1}\otimes \underline{3}'&=&\underline{3}'(11,12,13),\hs
\underline{1}'\otimes \underline{3}'=\underline{3}(11,12,13),\\
\underline{2} \otimes \underline{2} &=& \underline{1}(12+21)
\oplus \underline{1}'(12-21) \oplus \underline{2}(22,11),\\
\underline{2}\otimes \underline{3}&=&
\underline{3}\left((1+2)1,\om (1+\om 2)2,\om^2 (1+\om^2 2)
3\right)\crn && \oplus \underline{3}'\left((1-2)1,\om (1-\om
2)2,\om^2 (1-\om^2 2) 3\right) \\ \underline{2}\otimes
\underline{3}'&=& \underline{3}'\left((1+2)1,\om (1+\om 2)2,\om^2
(1+\om^2 2) 3\right)\crn &&\oplus \underline{3}\left((1-2)1,\om
(1-\om 2)2,\om^2 (1-\om^2 2) 3\right),\\
\underline{3} \otimes \underline{3} &=& \underline{1}(11+22+33)
\oplus \underline{2}(11+\om^2 22+ \om 33,11+\om 22+ \om^2 33) \crn
&&\oplus \underline{3}_s
(23+32,31+13,12+21)\oplus \underline{3}'_a(23-32,31-13,12-21),\\
 \underline{3}' \otimes
\underline{3}' &=&\underline{1}(11+22+33) \oplus
\underline{2}(11+\om^2 22+ \om 33,11+\om 22+ \om^2 33) \crn
&&\oplus \underline{3}_s
(23+32,31+13,12+21)\oplus \underline{3}'_a(23-32,31-13,12-21),\\
\underline{3} \otimes \underline{3}' &=&\underline{1}'(11+22+33)
\oplus \underline{2}(11+\om^2 22+ \om 33,-11-\om 22-\om^2 33) \crn
&&\oplus \underline{3}'_s (23+32,31+13,12+21)\oplus
\underline{3}_a(23-32,31-13,12-21),\eea where the subscripts $s$
and $a$ respectively refer to their symmetric and antisymmetric
product combinations as explicitly pointed out. We also notice
that many group multiplication rules above have similar forms as
those of $S_3$ and $A_4$ groups \cite{A4,kj}.

In the text we usually use the following notations, for example,
$(xy')_{\underline{3}}=
[xy']_{\underline{3}}\equiv(x_2y'_3-x_3y'_2,x_3y'_1-x_1y'_3,x_1y'_2-x_2y'_1)$
which is the Clebsch-Gordan coefficients of $\underline{3}_a$ in
the decomposition of $\underline{3}\otimes \underline{3}'$, where
as mentioned $x=(x_1,x_2,x_3)\sim \underline{3}$ and
$y'=(y'_1,y'_2,y'_3)\sim \underline{3}'$.

The rules to conjugate the representations \underline{1},
\underline{1}$'$, \underline{2}, \underline{3}, and
\underline{3}$'$ are given by \bea
\underline{2}^*(1^*,2^*)&=&\underline{2}(2^*,1^*),\hs
\underline{1}^*(1^*)=\underline{1}(1^*),\hs
\underline{1}'^*(1^*)=\underline{1}'(1^*),\\
\underline{3}^*(1^*,2^*,3^*)&=&\underline{3}(1^*,2^*,3^*),\hs
\underline{3}'^*(1^*,2^*,3^*)=\underline{3}'(1^*,2^*,3^*),\eea
where, for example, $\underline{2}^*(1^*,2^*)$ denotes some
$\underline{2}^*$ multiplet of the form $(x^*_1,x^*_2)\sim
\underline{2}^*$.

\section{\label{apt}The numbers}

In the following we will explicitly point out the lepton number
($L$) and lepton parity ($P_l$) of the model particles (notice
that the family indices are suppressed): \bc
\begin{tabular}{|c|c|c|}
  \hline
Particles & $L$ & $P_l$  \\ \hline
 $N_R$, $u$, $d$,  $\phi^+_1$,$\phi'^+_1$, $\phi^0_2$,$\phi'^0_2$, $\eta^0_1$,$\eta'^0_1$, $\eta^-_2$,$\eta'^-_2$
  $\chi^0_3$, $\sigma^0_{33}$, $s^0_{33}$ & 0 & 1  \\ \hline
  $\nu_L$, $l$, $U$, $D^*$, $\phi^+_3$,$\phi'^+_3$, $\eta^0_3$,$\eta'^0_3$, $\chi^{0*}_1$, $\chi^+_2$,
   $\sigma^0_{13}$,
   $\sigma^+_{23}$, $s^0_{13}$, $s^+_{23}$ & $-1$ & $-1$
   \\ \hline
   $\sigma^{0}_{11}$, $\sigma^{+}_{12}$, $\sigma^{++}_{22}$,
   $s^{0}_{11}$, $s^{+}_{12}$, $s^{++}_{22}$ & $-2$ & 1 \\ \hline
\end{tabular}\ec

\end{document}